\documentclass[manuscript,screen,nonacm]{acmart}

\AtBeginDocument{%
  }

\usepackage{subfig}

\newcommand{\toolname}{ReDSEa}

\begin{document}

\title{
Acceleration of Data Analytics on Heterogeneous
Supercloud Systems}

\author{Georgios Zacharopoulos}
\authornote{Both authors contributed equally to this research.}
\email{georgios.zacharopoulos@huawei.com}
\orcid{0000-0002-6644-5200}
\affiliation{%
  \institution{Computing Systems Lab, Huawei Research, Zurich}
  \country{Switzerland}
}

\author{Ilias Bournias}
\authornotemark[1]
\affiliation{%
  \institution{Computing Systems Lab, Huawei Research, Zurich}
  \country{Switzerland}
}
\email{ilias.bournias@huawei.com}
\orcid{https://orcid.org/0009-0000-5025-7988}

\author{Lukas Cavigelli}
\affiliation{%
  \institution{Computing Systems Lab, Huawei Research, Zurich}
  \country{Switzerland}
}
\email{lukas.cavigelli@huawei.com}
\orcid{0000-0003-1767-7715}

\renewcommand{\shortauthors}{Zacharopoulos et al.}

\begin{abstract}

Heterogeneous Supercloud systems are transforming data analytics by enabling scalable and efficient task distribution across diverse resources. This paper presents computational models and performance estimation techniques tailored for accelerating Dense Cholesky (CH) Decomposition and Singular Value Decomposition (SVD)-based analytics on Heterogeneous Superclouds.
Our \toolname\ tool-chain automates mapping, load balancing, scheduling, parallelism, and computation overlap.
Implemented on a heterogeneous system with a Huawei Kunpeng 920 ARM CPU and an Ascend 910 AI accelerator, our LLVM compiler tool-chain employs novel performance models for recursive, iterative, and blocked computations, achieving up to 17x speedup for CH and 88x for SVD over fully optimized 48-core CPU implementations.
\end{abstract}

\maketitle

\section{Introduction}

Heterogeneous computing has emerged as a vital paradigm in modern computing, leveraging a combination of diverse processor types and architectures to enhance performance, power efficiency, and cost-effectiveness. The concept of the Supercloud, introduced by McColl et al., represents an advanced heterogeneous cloud architecture specifically engineered to meet the demanding processing needs of AI, big data, and high-performance computing (HPC) applications. Within Supercloud systems, the integration of CPUs with specialized accelerators, such as GPUs, TPUs, and DPUs, capitalizes on the unique strengths of each component, fostering a synergy that significantly boosts performance and energy efficiency.

In these large-scale heterogeneous computing environments, high-speed interconnects enable different processors to collaborate efficiently through software that distributes and coordinates workloads at various levels of granularity. This ranges from low-level hardware acceleration to high-level task scheduling and load balancing. Such systems are particularly effective in parallel processing tasks vital for advanced simulations and machine learning algorithms, offering opportunities for various forms of parallel execution, including Instruction Level (ILP), Task Level (TLP), and Pipeline (PP) Parallelism.

However, designing complex Supercloud heterogeneous systems poses significant challenges that are both time-consuming and demanding. Engineers must carefully consider physical resource constraints, communication costs, and other factors when deciding which portions of an application should be accelerated on GPUs or hardware accelerators and which should run on general-purpose CPUs. Applications in domains like Deep Learning, Extended Reality (XR), and Autonomous Vehicles benefit from a Design Space Exploration (DSE) methodology based on performance estimation models. This methodology can (a) utilize all forms of parallelism, (b) be driven by the application source code or a graph representation of a computation model, and (c) automatically determine which parts of the application should be hardware-accelerated.

Supervised and unsupervised learning are fundamental to contemporary machine learning, playing vital roles in processing and interpreting complex data sets. A key strategy to enhance training efficiency for both types of learning is reducing features and dimensions. Singular Value Decomposition (SVD) is pivotal in executing Principal Component Analysis (PCA) for feature and dimensionality reduction, highlighting the most informative features and compressing the data.
However, traditional SVD algorithms are limited by memory bandwidth, impacting their performance. Therefore, novel methods have been proposed to accelerate computations by offloading them to GPUs or accelerators.

In our previous work~\cite{Zacharopoulos2023ReDSEa}, we introduced the \toolname\ tool-chain and applied it to the Triangular System Solver (TS). In this paper, we extend the methodology to Dense Cholesky Factorization and SVD, which require new performance models.

Our contributions are as follows:

\begin{itemize}

\item We explore three distinct computation models (recursive, iterative and blocked) for mapping Dense Cholesky Factorization and SVD to Supercloud heterogeneous systems.

\item We introduce novel models of performance estimation (computation and communication) for every computation model along with load balancing, parallelism and scheduling optimizations to maximize performance.

\item We extend \toolname~\cite{Zacharopoulos2023ReDSEa}, our LLVM-based automated tool, to utilize these models of performance estimation and select tasks as candidates for hardware acceleration.

\item We evaluate all three models of computation (recursive, iterative and blocked), and their  optimizations, using Dense Cholesky Factorization and SVD on an ARM 48-core CPU \cite{kunpeng} and an Ascend 910 \cite{ascend910} AI HW accelerator.

\end{itemize}

\section{Related Work}
\label{sec:rw}

The use of cloud heterogeneous systems to accelerate Linear Algebra has gained significant attention in research during recent years \cite{DONGARRA2017General}. Monil \textit{et al.} \cite{Lara2022Laris} introduce LaRIS, a portable framework for LAPACK functionalities on Heterogeneous Systems. LaRIS uses IRIS \cite{Iris2021Kim} to dynamically select the vendor-library kernel and suitable processor architecture at run-time, making the orchestration simpler when different architectures coexist in a single node. Valero-Lara et al. \cite{LaraTrsm} evaluate the use of OpenMP tasking with target GPU offloading as a potential solution for programming productivity and performance on heterogeneous systems. As a test case, the Triangular Solver (TS) routine is used.  In \cite{Miniskar2022}, the IRIS-BLAS framework is presented, which is built on top of IRIS. It has the capability to utilize all the available architectures on a heterogeneous system, using the appropriate BLAS library based on the task mapping at run time. Thus, the application developers are relieved from modifying the application source code. All three works mentioned previously follow the iterative and blocking approach.

In \cite{Relapack}, the authors introduce ReLAPACK, a methodology that makes use of recursive algorithms in order to implement LAPACK functionalities. They claim that, contrary to blocked algorithms, recursive algorithms do not require significant tuning to define a proper granularity/refinement level. They make this assumption for shared memory architectures. However, in all the aforementioned papers, there is not a model to determine both the refinement level and the expected performance.

A theoretical model, which predicts the refinement level and performance of Linear Algebra acceleration in homogeneous systems, is analyzed in \cite{beaumont2020TheoreticalCholesky}. Their model is based on the number of operations of kernels that are used in their framework, such as TRSM, SYRK and GEMM. The application that they target is the Dense Cholesky Factorization. Nevertheless, communication costs
in a homogeneous system are not taken into account while simulation results, and not experimental results, are presented. A similar approach for the same application is followed in \cite{BeaumontHetero}.

Recent years have seen increased interest in leveraging heterogeneous systems for accelerating SVD computations~\cite{gates2018accelerating,Liu2013Secularheterogeneous}. Typically, these approaches divide the SVD process into two stages, optimizing the use of heterogeneous resources.

The initial task involves transforming a full matrix into a bidiagonal form. Faverge \textit{et al.}~\cite{Faverge2017Bidag} investigate tiled algorithms for this transformation, focusing on orthogonal transformations and providing a critical path analysis for each algorithm. They demonstrate that R-BIDIAG exhibits a shorter critical path length compared to BIDIAG for tall and skinny matrices, and GREEDY-based schemes outperform other state of the art algorithms when resources are unlimited. They evaluate their approaches by conducting experiments on both a single multicore node and on a parallel distributed shared-memory system consisting of a few multicore nodes. They conclude that QR and LQ reduction trees cannot overlap in their processes. Ltaief \textit{et al.}~\cite{2013LtaiefBidiagonal} explore the influence of tile size on execution time, linking it to the matrix bandwidth size post-reduction. They use a brute force mechanism to retrieve the optimal tile size of a specific problem size. It is also pointed out that the optimal tile size is not the same for every sub-task of the problem. Gates \textit{et al.}~\cite{gates2018accelerating} develop a 2-stage reduction algorithm to transform a full matrix to a bidiagonal one, accelerating the process by offloading matrix multiplications to GPUs and overlapping them with LQ and QR computations on the CPU. However, these studies lack automated tools for performance estimation based on hardware resources and granularity level to achieve maximum throughput.

The second stage focuses on the SVD of the bidiagonal matrix. Gates \textit{et al.}~\cite{gates2018accelerating} use the Divide and Conquer (D\&C) algorithm, achieving better performance compared to QR factorization methods~\cite{2009LahabarQRinsteadofDC}. They accelerate matrix multiplications with GPUs, while the CPU handles secular equation solving and singular vector computation. Liu \textit{et al.}~\cite{Liu2013Secularheterogeneous} also use the Divide and Conquer approach but on the top of that, they offload the secular equation computation to accelerators. They examine the performance of different algorithms for this task and they try to reduce the number of the floating point operations of the different algorithms in order to decrease latency. However, these papers do not delve into the recursion depth of the Divide and Conquer algorithm, a crucial aspect in heterogeneous systems.

Additionally, compiler-based methodologies have been introduced for automatic performance estimation in heterogeneous systems \cite{ZacharopoulosNov19, ZacharopoulosApr19}, focusing on general approaches using LLVM static analysis and dynamic profiling \cite{ZacharopoulosMar17}, or targeting specific parallelism types \cite{zacharopoulos23trireme}. However, they do not address diverse computation models which can be used to enhance performance and they focus mainly on streaming applications.

Design space exploration (DSE) methods have also been explored \cite{ferretti2022graph, BrumarEarlyDSE,2021Iliasiscas}. They focus on strategies for synthesizing accelerators and implementing optimizations in terms of performance improvement, accuracy, and reducing the area required for hardware implementations. Nevertheless, they target mainly FPGA devices and lack a potentially wider variety of hardware.

\section{Compiler Tool-chain}
\label{sec:compiler}

To address the challenges mentioned in the previous section, we build on
\textbf{\toolname} (\underline{Re}cursive \underline{D}esign \underline{S}pace \underline{E}xploration \underline{A}utomation):
the tool-chain we introduced in~\cite{Zacharopoulos2023ReDSEa}, based on the LLVM \cite{LattnerMar04} Compiler Infrastructure (Version 14.0.0).

\subsection{Compiler Analysis}

The first stage of our methodology is analyzing the application that is going to be mapped to a heterogeneous system or cloud architecture. The analysis phase includes an automated process that provides our compiler infrastructure with necessary information so as to guide the subsequent steps (Cost Models and Design Space Exploration) of the automation methodology. It can be viewed as the \textbf{input} to our automated system.
The analysis is performed by generating the respective LLVM-IR from the application source code (typically written in C, C++). A number of LLVM analysis passes, developed within the scope of \toolname, analyze the intermediate representation of the applications and estimate the latency due to computation of every potential task, or otherwise described, of every node of the data flow graph. Furthermore, the communication cost is estimated by extracting the amount of data that is read and stored by every task. The data requirements, along with the available bandwidth of a target architecture, allows for an estimation of the latency, due to communication.

\subsection{Cost Models}

To obtain an early evaluation of the potential performance of every computational node, we need to introduce evaluation/cost models that can perform estimations of both computation and communication latency for the components of a Supercloud heterogeneous system, such as CPUs, GPUs, HW accelerators, etc.

To estimate the latency of a computational node that is mapped to multiple CPUs, GPUs and/or HW accelerators, we need to estimate
\begin{enumerate}
    \item the latency of the computation on the CPUs,
    \item the critical path of the computation that is offloaded to GPUs and/or HW accelerators,
    \item the latency to transfer the data from the host main memory to the HW accelerator (Host-to-Device, H2D) and vice versa (Device-to-Host, D2H), and
    \item the synchronization/invocation overhead.
\end{enumerate}
Latency = CPU Computation + HW Computation + Communication + Synchronization/Invocation.

Let $S = \{ S_1, S_2, \ldots, S_N \}$ be a set of nodes (tasks), with associated HW computation latency $t_{HWcomp}^{(i)}$, HW communication latency $t_{HWcomm}^{(i)}$ and synchronization/invocation overhead. For every node $i$, the cumulative latency will be $t_{HW}^{(i)} = t_{HWcomp}^{(i)} + t_{HWcomm}^{(i)} + t_{synch}^{(i)}\ |\ i=1,\dots, N$.

\subsection{Design Space Exploration}
Based on the cost models of the previous subsection, a list of candidates for acceleration is generated.
The selection (branch-and-bound) algorithm recursively explores the subsets of the list of candidates, in a similar manner to the Bron-Kerbosch algorithm \cite{BronKerbosch73}. The output returned is the set
with the highest speedup (minimum cumulative latency) that stays within the user defined resource budget, which is translated as the amount of resources available for hardware acceleration.

\section{Linear Algebra Applications}
\label{sec:linalg}

We have selected two representative instances from the linear algebra domain to use as our test cases: Dense Cholesky Factorization and Singular Value Decomposition (SVD) from LAPACK \cite{dongarabooklapack1999}.

\begin{figure}[t]
\centering
\includegraphics[width=0.8\linewidth]{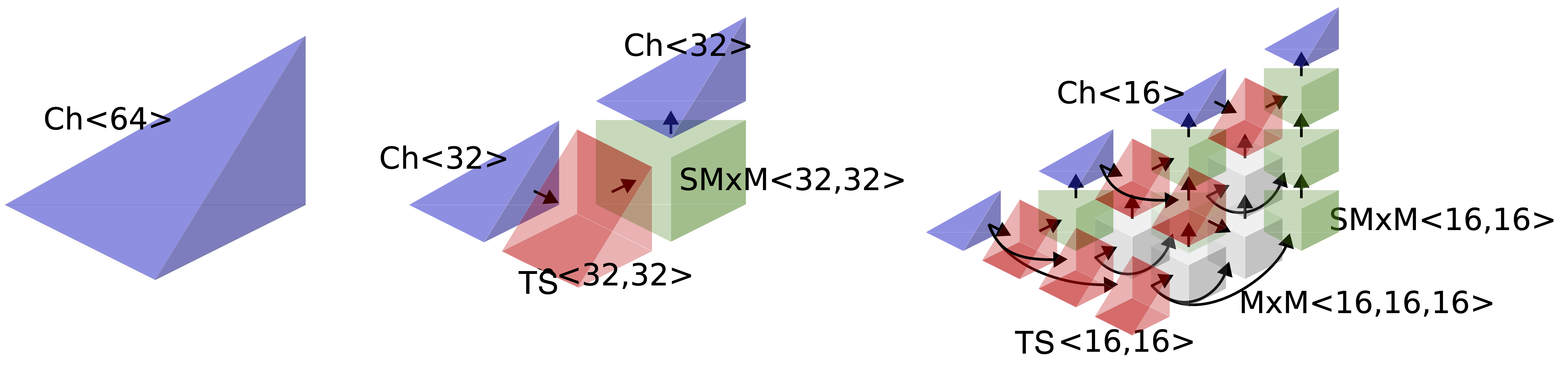}
\vspace{-0.2cm}
\caption{Data Flow Graph (DFG) of the recursive implementation of dense Cholesky Factorization \texttt{CH<n>} and its refined (decomposed) nodes \texttt{CH<n/2>}, triangular system solver \texttt{TS<n/2>} and the general matrix multiplication \texttt{gemm<n/2,n/2>}.
}
\vspace{-0.5cm}
\label{fig:cholecky-rec}
\end{figure}

\subsection{Dense Cholesky Factorization}
Cholesky Factorization (Figure \ref{fig:cholecky-rec}), also known as Cholesky Decomposition, is a numerical algorithm used in linear algebra to decompose a Hermitian, positive-definite matrix into a lower triangular matrix and its transpose. This decomposition is useful in many applications, such as solving systems of linear equations, computing eigenvalues and eigenvectors, and simulating Gaussian random variables. It has numerous applications in the fields of numerical optimization, computational fluid dynamics, structural analysis, and Bayesian statistics.

Given a dense $n \times n$ positive-definite real symmetric matrix $A$, the problem is to compute the unique lower-triangular $n \times n$ real matrix $L$ with positive diagonal entries such that $A = LL^T$.
The matrices $A$ and $L$ are blocked in a similar manner, divided into sub-matrices up, mid, and low corresponding to the upper left, lower left, and lower right blocks of $L$ and $A$, so that:
\begin{align*}
\small
\begin{pmatrix}
A.up & A.mid^T\\
A.mid & A.low
 \end{pmatrix}
 =
 \begin{pmatrix}
 L.up & 0 \\
 L.mid & L.low
 \end{pmatrix}
 \begin{pmatrix}
 L.up^T & L.mid^T\\
 0 & L.low^T
\end{pmatrix}
\end{align*}
Thus, $A.up = L.up \cdot L.up^T$, $A.mid = L.mid \cdot L.up^T$, and $A.low = L.mid \cdot L.mid^T + L.low \cdot L.low^T$.
Hence, the
Cholesky factorization of $A$ can be computed in four steps as follows:

\begin{enumerate}
\item Compute the Cholesky factorization of $A.up$ to obtain $L.up$
\item Compute $L.mid = A.mid(L.up^T)^{-1}$  by solving a triangular system (TS) with multiple right-hand sides
\item Update $A.low = A.low - L.mid\cdot L.mid^T$ (gemm)
\item Compute Cholesky factorization of $A.low$ to obtain $L.low$
\end{enumerate}

\subsection{Accelerated SVD for Heterogeneous Systems}
\label{sec:svd}

\begin{figure}[t]
\centering
\includegraphics[width=0.6\linewidth]{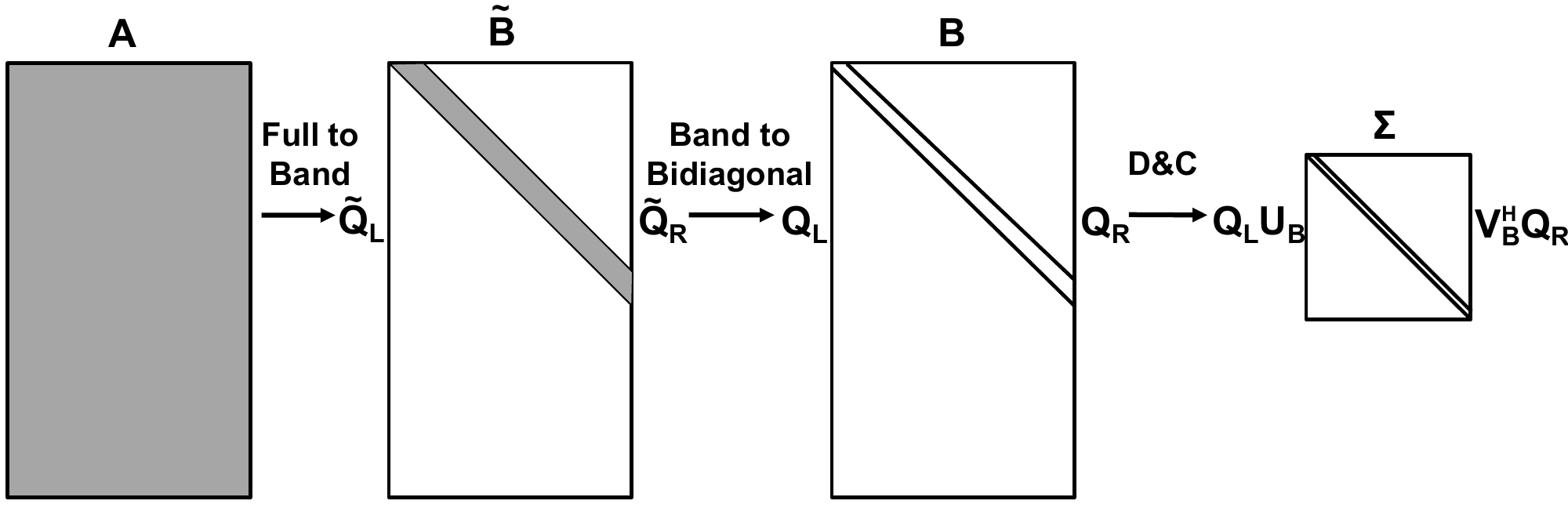}
\vspace{-0.3cm}
\caption{2-stage reduction to bidiagonal form followed by Divide and Conquer (D\&C) SVD.
}
\vspace{-0.5cm}
\label{fig:twostage}
\end{figure}
The traditional SVD algorithm employs a \emph{1-stage} reduction to bidiagonal form, relying on matrix-vector multiplications. This approach is often hindered by limited memory bandwidth by underutilizing the available computation resources, adversely affecting performance. To address this issue, a \emph{2-stage} reduction algorithm to bidiagonal form which makes use of matrix-matrix multiplications has been proposed \cite{gates2018accelerating}, among others. As illustrated in Figure~\ref{fig:twostage}, this algorithm initially reduces the matrix to a band form utilizing high-performance Level 3 BLAS functions, followed by reducing the band matrix to bidiagonal form using cache-optimized kernels with dynamic scheduling. This strategy mitigates memory bandwidth limitations, reduces communication and synchronization latency, and enhances computational workload.

The SVD of an $m \times n$ matrix $A$ is defined as $A = U\Sigma V^H$. Here, $U$ and $V$ are unitary matrices, and $\Sigma$ is a real diagonal matrix with non-negative elements $\sigma_i$, which represent the singular values of $A$. The first $\min(m, n)$ columns of $U$ and $V$ constitute the left and right singular vectors of $A$, respectively. The Golub–Kahan–Reinsch algorithm \cite{golub1965calculating, golub1971singular} performs SVD in three steps:

\begin{enumerate}
    \item Bidiagonalization: Transform  $A$ into a bidiagonal matrix $B$ such that $A=Q_L B Q_R$ with $Q_L,Q_R$ unitary matrices,
    \item Bidiagonal SVD: Employ QR or Divide and Conquer methods~\cite{gu1995divide} to obtain $B=U_B \Sigma V_B^H$,
    \item Extraction of singular vectors through back-transformation, $A=Q_L B Q_R = Q_L U_B \Sigma V_B^H Q_R = U \Sigma V^H$.
\end{enumerate}

\subsubsection{2-Stage Bidiagonalization}
The 2-stage reduction process begins by converting the full matrix into an upper band form, then further reducing it to upper bidiagonal. Specifically,
\begin{align*}
    A=\tilde{Q}_L\tilde{B}\tilde{Q}_R=\tilde{Q}_L \hat{Q}_L B \hat{Q}_R \tilde{Q}_R = Q_L B Q_R,
\end{align*}
with all matrices $Q$ unitary, $\tilde{B}$ the upper band matrix, and $B$ the upper bidiagonal matrix.

{\textbf{Full-to-Band.}}
The first phase involves transforming the matrix to band form, crucially removing data dependencies with the trailing matrix during panel factorization. This is key to overcoming the bottleneck caused by matrix-vector operations in the panel.

\begin{figure}[t]
\centering
\includegraphics[width=0.7\linewidth]{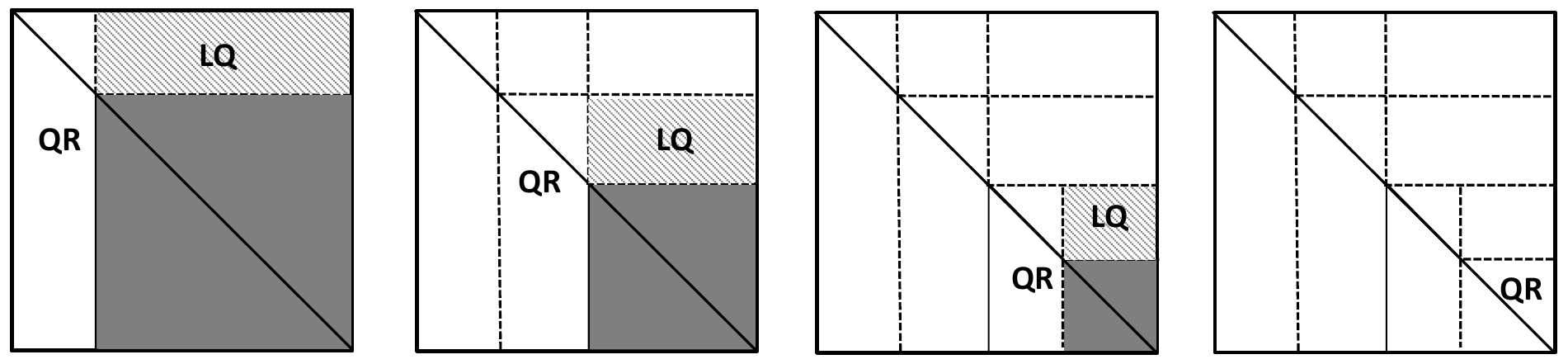}
\vspace{-0.4cm}
\caption{Bidiagonalization steps using 2-stage reduction. Each grey square represents the area of gemm computation that is offloaded to an accelerator. The white rectangles of QR, as well as LQ, reside on the host CPU.}
\vspace{-0.5cm}
\label{fig:svd-iter-unfold}
\end{figure}

The procedure involves a QR panel factorization of a block column to remove entries below the diagonal, followed by an update of the trailing matrix. This is succeeded by an LQ panel factorization of a block row to eliminate entries right of the upper band width of the matrix, and then another update of the trailing matrix. Figure~\ref{fig:svd-iter-unfold} illustrates this process, with the trailing matrix updates shown as dark grey rectangles. These updates are offloaded to accelerators, while the QR and LQ panels are computed on the host CPU.

{\textbf{Band-to-Bidiagonal.}}
In the second stage, the upper band matrix is further reduced to upper bidiagonal form. Due to the limited parallelism, memory bandwidth constraints, and the need for CPU cache optimizations, using accelerators for this stage offers only marginal benefits. Hence, this stage's computations are primarily performed on the host CPU. It is important to note that the bulk of the computational work occurs in the initial stage (from full to band form), which consequently lessens the computational load in this later stage.

\subsubsection{Divide and Conquer (D\&C)} \label{divide and conquer theory}
The Divide and Conquer algorithm is employed for computing the SVD of a bidiagonal matrix $B$ after its transformation into bidiagonal form. The matrix $B$ is divided into two bidiagonal submatrices, $B_{1}$ and $B_{2}$, as illustrated below ($e_{k}$ is the $k$-th column of an identity matrix). On each sub-matrix, the SVD is computed separately:
\begin{align*}
B=
\begin{pmatrix}
B_{1} & \alpha_{k}e_{k} & 0\\
0 & \beta_{k}e_{1} & B_{2}
\end{pmatrix}, \quad
B_{i}=\begin{pmatrix}Q_{i} & q_{i}\end{pmatrix}\begin{pmatrix} \Sigma_{i} \\ 0\end{pmatrix}W_{i}^H.
\end{align*}
By substituting $B_{1}$ and $B_{2}$ with their SVD decompositions and implementing Givens rotations~\cite{gates2018accelerating} ($r_{0}$, $c_{0}$, $s_{0}$), we obtain $B=Q\overline{M}W^H$. This results in
\begin{align*}
Q&=\begin{pmatrix}
  c_{0}q_{1} & Q_{1}    & 0     & -s_{0}q_{1} \\
  s_{0}l_{1} & 0        & Q_{2} & c_{0}q_{2} \\
\end{pmatrix}, \\
\overline{M}&=\begin{pmatrix} \tilde{M} \\ 0 \end{pmatrix}
=\begin{pmatrix}
  r_{0} & 0 & 0 \\
  \alpha_{k}l_{1} & D_{1} & 0 \\
  \beta_{k}f_{2} & 0 & D_{2} \\
  0 & 0 & 0
\end{pmatrix}, \\
W^{H}&=\begin{pmatrix}
  0 & W_{1} & 0 \\
  1 & 0 & 0 \\
  0 & 0 & W_{2} \\
\end{pmatrix}.
\end{align*}

If $\tilde{M}=U_{\tilde{M}}\Sigma_{\tilde{M}} V^{H}_{\tilde{M}}$ is the SVD of $\tilde{M}$, then $B=\begin{pmatrix}QU_{\tilde{M}}\end{pmatrix}\begin{pmatrix}\Sigma_{\tilde{M}}\\0\end{pmatrix}\begin{pmatrix}WV_{\tilde{M}}\end{pmatrix}^{H}$. The singular values of $B$ match those of $\overline{M}$. To determine the singular values of $\overline{M}$, we adopt the methods proposed by \cite{gates2018accelerating} and \cite{Liu2013Secularheterogeneous}, solving the secular equation formed by the elements of $\overline{M}$. After computing the singular values, the singular vectors $(U_{\tilde{M}},V_{\tilde{M}}^{H})$ are derived as per~\cite{gates2018accelerating} and multiplied with $Q$ and $W^{H}$, respectively, to yield the singular vectors of $B$ (U$_{B}$,V$_{B}^{H}$). These multiplications are offloaded to the accelerators, while the computations for singular values and vectors, and the computation of $\overline{M}$ remain on the host, utilizing LAPACK subroutines.

The Divide and Conquer algorithm is recursively applied to $B_{1}$ and $B_{2}$ and their subsequent sub-matrices. However, at a certain recursion depth, the Divide and Conquer algorithm becomes less beneficial due to excessively fine granularity, leading to the employment of a classic QR SVD approach for bidiagonal matrices~\cite{gates2018accelerating}.

\subsubsection{Back-Transformation of Singular Vectors}
\label{sec:bt}

It is important to note that increased computational effort in the full-to-band reduction leads to more intensive computation in the first back-transformation stage due to the increased number of Householder reflectors. All back-transformation steps are offloaded to accelerators as they primarily involve matrix-matrix multiplications.

\section{Models of Computation}
\label{sec:models}

Three models of computation are explored:
recursive, iterative, and blocked \cite{Relapack,DONGARRA2017General,Lara2022Laris,LaraTrsm}.

\subsection{Recursive}
\label{sec-rec}

The recursive model of computation offers a decomposition of the initial problem to several tasks, as shown in Figure \ref{fig:cholecky-rec}. It also provides the opportunity to expose parallelism and finally to explore various levels of granularity in order to make the best architectural decisions and use efficiently any given available software (general purpose CPUs) and hardware accelerators resources.

\begin{figure}[t]
\centering
\includegraphics[width=0.55\linewidth]{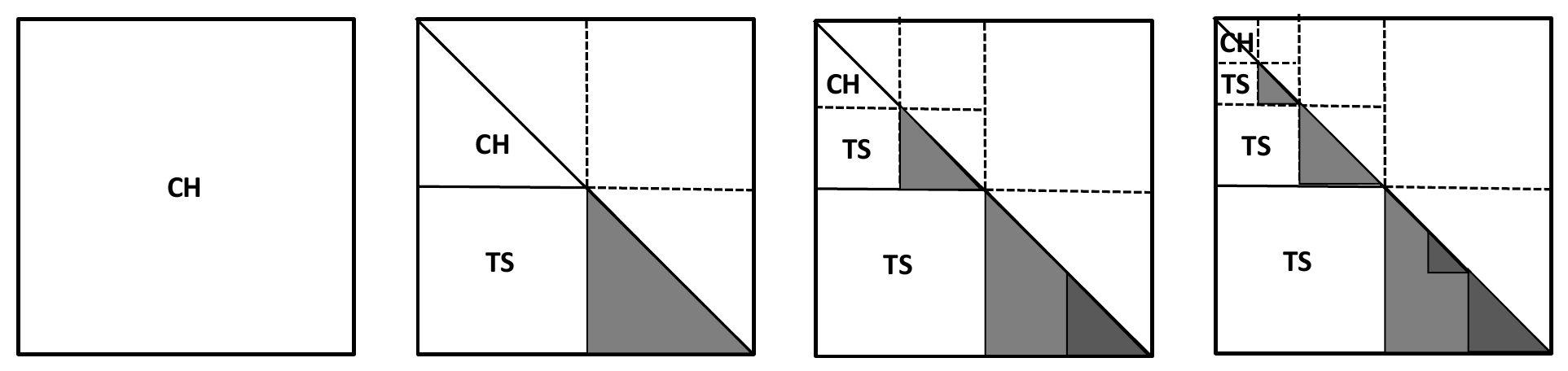}
\includegraphics[width=0.55\linewidth]{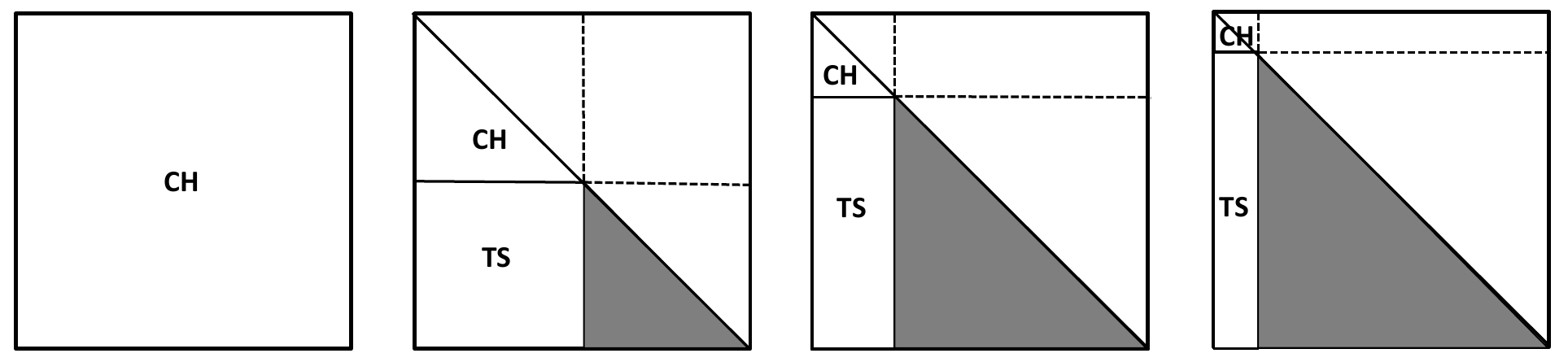}
\includegraphics[width=0.55\linewidth]{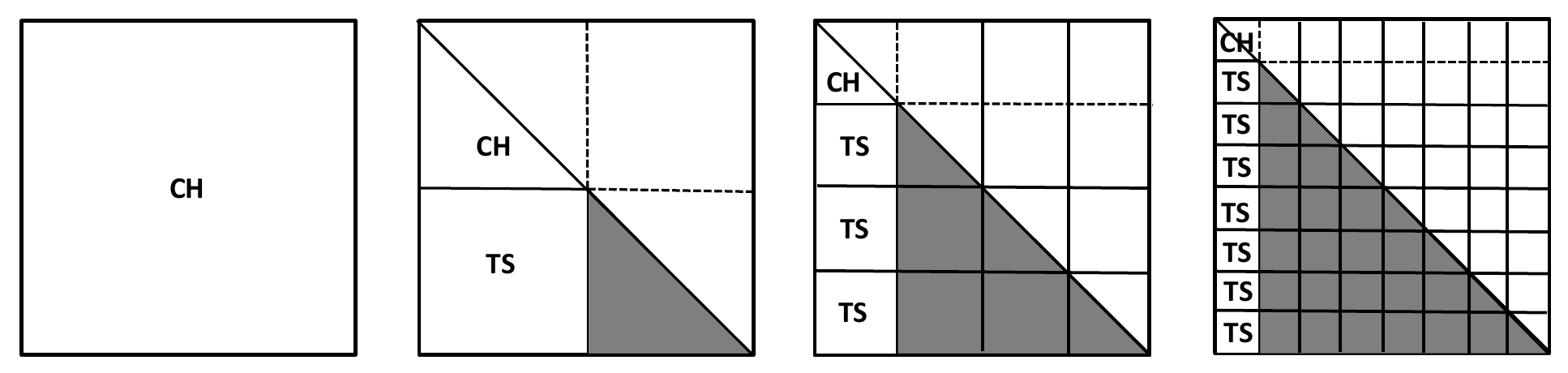}
\vspace{-0.4cm}
\caption{Recursive (top), Iterative (middle) and Blocked (bottom) models respectively for Cholesky (Iteration: 0, 1, 2 and 3 Refinement Level: 1, 2, 4 and 8 respectively).
Each grey half-square (square triangle) represents the gemm computation that is offloaded to an accelerator. The annotated white square triangles (Cholesky) and rectangles (TS) represent the part of the computation that remains in the host CPU.}
\label{fig:cholesky-models}
\end{figure}
\subsubsection {\textbf{Dense Cholesky Factorization}}
In the case of Cholesky Factorization (CH), as depicted in Figure \ref{fig:cholesky-models} (top), the initial problem of size $n$ is decomposed to the tasks: \texttt{CH<n/2>}, triangular system \texttt{TS<n/2>}, and general matrix-matrix multiplication \texttt{gemm<n/2,n/2>}, along with the respective dependencies. We keep the CH and TS tasks in the host CPU and we offload the gemm parts of the computation to accelerators, taking advantage of its exposed data level parallelism as we decomposed the initial CH problem.

We notice that as we refine further the CH tasks, more gemm tasks become available and, hence, greater part of the computation can be accelerated.
The next refinement stage is applied \emph{only} if it benefits performance, i.e., if the following \textbf{condition} is satisfied:
\begin{align*}
    2 \cdot t_{CH}^{(i+1)} < t_{CH}^{(i)}\ |\ i=0,\dots, N,
\end{align*}
where $t_{CH}^{(i)}$ is the latency of the task CH for iteration $i$.
We define as $r(i)$ the \textbf{Refinement Level} of iteration $i$:
\begin{align}
    r(i)=2^i\ |\ i=0,\dots, N.
\end{align}

Both TS and gemm tasks accumulate varied granularities from previous iterations as the iteration itself, and the respective refinement, increases.
The models describing the performance on a heterogeneous architecture are
\begin{align*}
    &t_{comp}^{(i)} = r(i) \cdot t_{CH}^{(i)} + \sum_{j=0}^{i-1} r(j) \cdot \left(t_{gemm}^{(j)} + t_{TS}^{(j)}\right), \\
    &t_{comm}^{(i)} = \sum_{j=0}^{i-1} r(j)\cdot t_{comm, H{2}D+D2H}^{(j)}.
\end{align*}
\subsubsection {\textbf{Divide and Conquer}}
The Divide and Conquer algorithm approach, distinct from the previous stages, employs a recursive method. As detailed in Section~\ref{divide and conquer theory}, each $B_{i}$ subproblem of $B$ is recursively divided into two $B^{'}_{i}$ subproblems of $B_{i}$, each independent of the other.

Gemm tasks, necessary for computing the singular vectors of $B$, are accelerated to enhance performance. The host CPU handles the computation of singular values and singular vectors of the middle matrix $\overline{M}$.

The next refinement stage $i+1$ in the recursive approach is contingent upon the following \textbf{condition}:
\begin{align*}
&2 \cdot t_{SVD}^{(i+1)} + t_{Sv}^{(i+1)}
    + t_{SV}^{(i+1)}   <     t_{SVD}^{(i)}
\end{align*}

where $t_{SVD}$ is the latency of the SVD of $B_{i}$, and $t_{Sv}$ and $t_{SV}$ are the latencies of the singular values and singular vectors computation of $\overline{M}$, respectively.

To estimate computation and communication cost for the recursive approach, the following models are given respectively:
\begin{align*}
    &t_{comp}^{(i)} = r(i) \cdot t^{(i)}_{SVD} + \sum_{j=0}^{i-1} r(j) \cdot \left(t_{Sv}^{(j)}+t_{SV}^{(j)}+4t_{gemm}^{(j)}\right),\\
    &t_{comm}^{(i)} = \sum_{j=0}^{i-1} 4 \cdot r(j) \cdot \left(t_{comm,H2D}^{(j)}+t_{comm,D2H}^{(j)}\right).
\end{align*}

\subsection{Iterative}
\label{sec-iter)}

Once the granularity of the computation that remains to the host has been determined, an iterative model of computation can be offered as an alternative, as seen in Figure \ref{fig:cholesky-models} (middle).\\
There are two main reasons to favor an iterative approach over a recursive one: a) It can offer better utilization of the accelerators that are available, as fewer accelerators are dedicated to compute relatively smaller parts of the computation compared to a recursive one, which gradually allocates smaller and smaller parts of the computation to be computed by the HW accelerators.
b) It demands less engineering effort to be implemented while at the same time the performance is not sacrificed, but instead it is kept equal, or slightly better, compared to a recursive one.
Furthermore, in the case of Dense Cholesky Factorization, the amount of computation that is offloaded to the accelerator is greater compared to the recursive model.

\begin{figure}[t]
\centering
\includegraphics[width=0.6\linewidth]{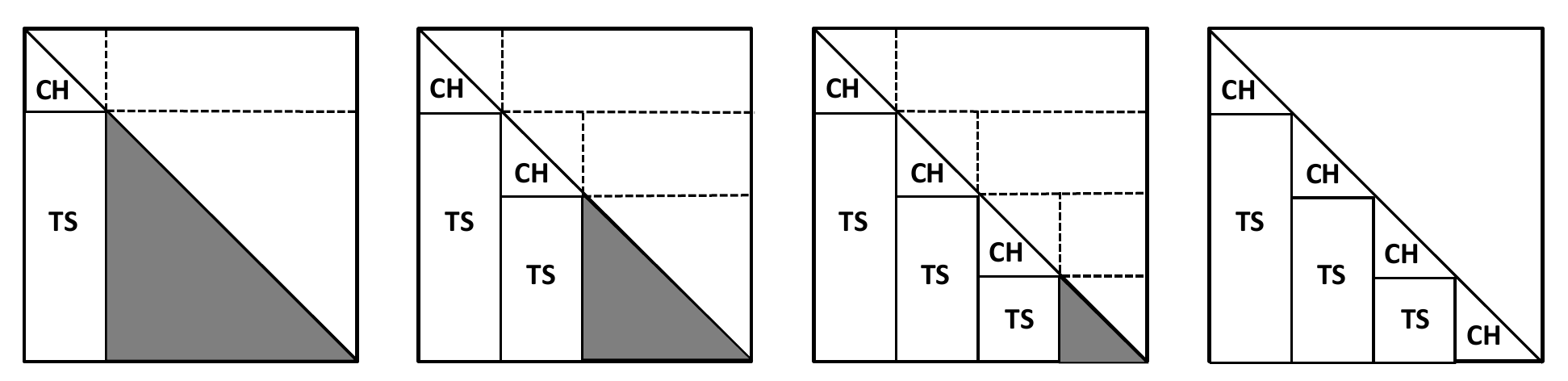}
\vspace{-0.5cm}
\caption{Iterative model for dense Cholesky Factorization (Iteration: 2 Refinement Level: 4).
Each grey half-square (square triangle) represents the accelerated gemm task. The rest of the computation remains in the host CPU. The gemm tasks vary in size across the four rounds of computation.
}
\vspace{-0.5cm}
\label{fig:chol-iter-time}
\end{figure}

\subsubsection{\textbf{Dense Cholesky Factorization}}
As shown in Figure~\ref{fig:cholesky-models}, in comparison to the recursive model, the amount of TS computation that resides at
 the host side is reduced, whereas the gemm computations that are available for acceleration are increased.

Since this is a more complex problem, it is necessary to show that the computation of the algorithm continues to one more dimension and it is fully unfolded in Figure \ref{fig:chol-iter-time}, for the example of $i=2$, $r(i)=4$ of the Cholesky iterative model. We showcase that there are four rounds of computation and three instances of gemm acceleration, which vary in size. The rounds of computation in the iterative approach are equal to the refinement level $r(i)$.

The respective models that estimate the expected
computation and communication latency of the iterative approach
for Cholesky Factorization are
\begin{align*}
    &t_{comp}^{(i)} = r(i) \cdot t_{CH}^{(i)} + \sum_{j=0}^{r(i)-2} t_{gemm}^{(i,j)} + t_{TS}^{(i,j)}, \\
    &t_{comm}^{(i)} = \sum_{j=0}^{r(i)-2} t_{comm,H2D+D2H}^{(j)}.
\end{align*}

\begin{figure}[t]
\centering
\includegraphics[width=0.6\linewidth]{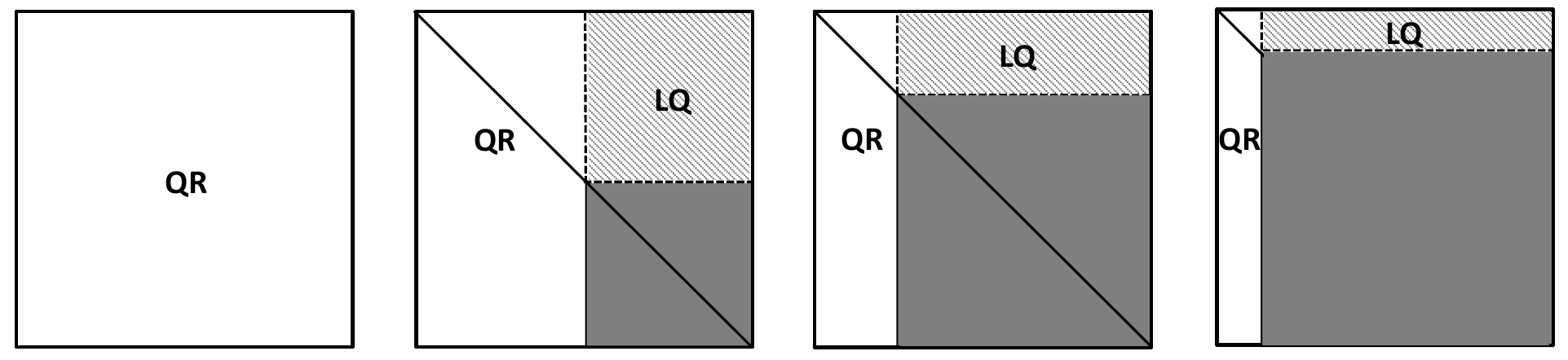}
\includegraphics[width=0.6\linewidth]{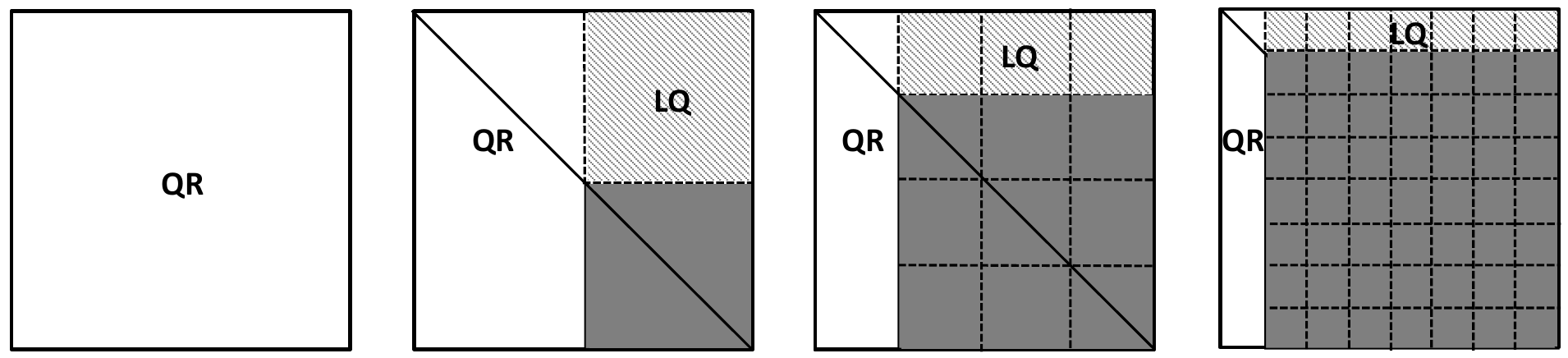}
\vspace{-0.4cm}
\caption{
Iterative (top) and Blocked (bottom) models for 2-stage SVD (Iteration: 0, 1, 2 and 3 Refinement Level: 1, 2, 4 and 8 respectively).
Each dark grey rectangle (square) represents the area of gemm computation that is offloaded to an accelerator. The white rectangles of QR and LQ represent the part of the computation that remains in the host CPU.
}
\vspace{-0.4cm}
\label{fig:svd-iter}
\end{figure}

\subsubsection{\textbf{Bidiagonalization}}
The iterative approach to 2-stage bidiagonalization is depicted in Figure~\ref{fig:svd-iter}. For clarity, we illustrate only the first panel of the bidiagonalization process at each iteration or level of \emph{refinement}. The concept of ``refinement'' here refers to a task's decomposition level, allowing for progressively finer granularity with each iteration.

We define the \textbf{Refinement Level} for iteration $i\in\{0,\dots,N\}$ as $r(i)=2^i$. For each iteration $i$ of the bidiagonalization decomposition/refinement, the dimensions of the initial $QR$ matrix $n \times n$ (at $i=0$) are halved from the size of the preceding iteration (thus, $n \times n/2$ for $i=1$, $n \times n/4$ for $i=2$, and so on).

The decision to proceed to the next refinement stage is based on a performance-related \textbf{condition}. This condition is applied \emph{only} if advancing to the next refinement level $i+1$ is projected to enhance performance. The following inequality quantitatively defines the criterion for proceeding with refinement in the iterative model of bidiagonalization:
\begin{align*}
    \sum_{j=0}^{r(i+1)-1} t_{QR}^{(i+1,j)} + t_{BD}^{(i+1)}   < \sum_{j=0}^{r(i)-1} t_{QR}^{(i,j)}  + t_{BD}^{(i)}
\end{align*}
where $t_{QR}^{(i,j)}$ represents the latency of the task QR for iteration $i$, and $t_{BD}^{(i)}$ is the latency of bidiagonalization.

For example, Figure~\ref{fig:svd-iter} demonstrates the computational equivalence for refinement levels 1 (no decomposition), 2, 4, and 8. With increasing refinement, the computational load of gemm tasks (depicted as dark grey rectangles) also increases, providing more opportunities for acceleration through offloading to available hardware resources.

Notably, the more computation is executed in the first stage (full-to-band), the less computational effort is needed in the second stage (band-to-bidiagonal), which mainly takes place on the host CPU and is not as amenable to acceleration. Therefore, our goal is to offload most of the computational load required for bidiagonalization to the first stage, thus expediting the overall process.

To estimate the computation and communication latencies in a heterogeneous Supercloud architecture, the following models are used:
\vspace{-0.2cm}
\begin{align*}
    &t_{comp}^{(i)} = \sum_{j=0}^{r(i)-1} t_{QR}^{(i,j)} +
    \sum_{j=0}^{r(i)-2} t_{LQ}^{(i,j)} +
    \sum_{j=0}^{r(i)-2} \sum_{k=0}^{N-1}   t_{gemm}^{(i,j,k)} , \\
    &t_{comm}^{(i)} = \sum_{j=0}^{r(i)-2} \sum_{k=0}^{N-1}   \left(t_{comm,H2D}^{(i,j,k)}+t_{comm,D2H}^{(i,j,k)}\right).
\end{align*}

\vspace{-0.5cm}
\subsection{Blocked}

\begin{figure}[t]
\centering
\includegraphics[width=0.5\linewidth]{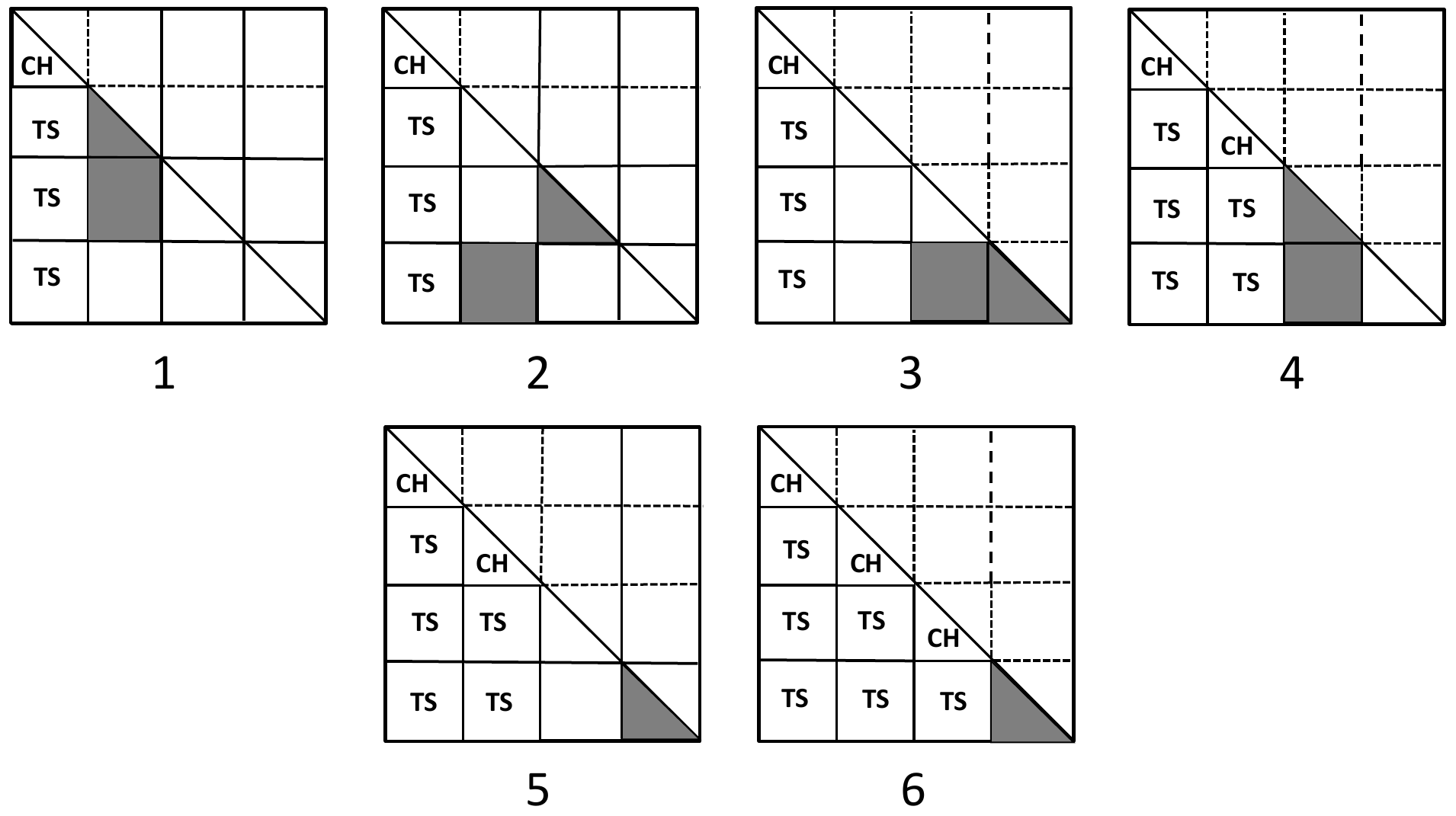}
\caption{Workflow in six rounds of computation for the blocked model of Dense Cholesky Factorization (refinement $r(2) = 4$, blocks $r(2)^2/8=2$).
}
\vspace{-0.5cm}
\label{fig:flow_ch}
\end{figure}

A blocked approach, as seen in Figure \ref{fig:cholesky-models} (bottom), can be used once the level of refinement and granularity of the computations that reside at the host CPU have been determined.
Several advantages emerge in the blocked approach, compared to a recursive or an iterative one.
These include a) better overall load balancing, b) more efficient use of HW accelerators resources, c) better scheduling and
d) potential task overlapping, aligning computation closer to the theoretical critical path for peak performance.

\subsubsection{Dense Cholesky Factorization}
The blocked model offers a computation in rounds for Dense Cholesky Factorization, as seen in Figure \ref{fig:flow_ch}. This workflow provides both better load balancing, compared to the previous models, and the opportunity for parallel and  overlapping computation of independent tasks.

For every iteration $i$ and a refinement level $r(i)$, the number of \textbf{rounds of computation} is
\begin{equation*}
\sum_{j=0}^{r(i)-2}
\left\lceil
\dfrac{ (r(i)-j-1) \cdot (r(i)-j)/2 }{r(i)^2/8}\cdot{\dfrac{r(i)^2}{8}}
\right\rceil,
\end{equation*}
where the number of \textbf{blocks per computation} is $r(i)^2/8$ per round and
the total number of blocks (gemm(i) computations) to be accelerated are
$\sum_{j=0}^{r(i)-2} (r(i)-j-1) \cdot (r(i)-j)/2$. In the example of Figure \ref{fig:flow_ch},
the number of blocks per round is $r(2)^2/8=2$,
the rounds of computation are $3 + 2 +1 = 6$ and
the total number of blocks is  $6+3+1=10$.

The  models  that compute  the expected  performance  and communication cost of the blocked approach for Dense Cholesky Factorization are
\begin{align*}
t_{comp}^{(i)} = &r(i) \cdot t_{CH}^{(i)} + (r(i) -1) \cdot \dfrac{r(i)}{2} \cdot t_{TS}^{(i)} + \\
          &\sum_{j=0}^{r(i)-2} (r(i)-j-1) \cdot \dfrac{(r(i)-j)}{2} \cdot t_{gemm}^{(i)}, \\
t_{comm}^{(i)} = &\sum_{j=0}^{r(i)-2} (r(i)-j-1) \cdot \dfrac{(r(i)-j)}{2} \cdot t_{comm, H2D+D2H}^{(i)}.
\end{align*}

\subsubsection{Bidiagonalization}

The computation and communication latency for a heterogeneous Supercloud in the blocked model are estimated as follows:
\begin{align*}
    t_{comp}^{(i)} = &\sum_{j=0}^{r(i)-1} t_{QR}^{(i,j)} + \sum_{j=0}^{r(i)-2} t_{LQ}^{(i,j)}\\
                     &+\sum_{j=0}^{r(i)-2} \left(r(i)-j-1\right)^2\cdot t_{gemm}^{(i,j)},\\
    t_{comm}^{(i)} = &\sum_{j=0}^{r(i)-2} \sum_{k=0}^{N-1} \left(t_{comm,H2D}^{(i,j,k)}+t_{comm,D2H}^{(i,j,k)}\right).
\end{align*}

\subsubsection{\textbf{Parallel Computation}}

The blocked approach offers the possibility to use multiple accelerators where equivalent loads of gemm(i) computations can be computed.
For example, a parallel implementation of the Cholesky case (Figure \ref{fig:flow_ch}) would require a minimum of $2$ available accelerators to conclude the computation in $6$ rounds. The respective model estimating the performance for a parallel blocked approach is the following:
\begin{align*}
t_{comp}^{(i)} &= r(i) \cdot t_{CH}^{(i)} +
(r(i) - 1) \cdot \dfrac{r(i)}{2} \cdot t_{TS}^{(i)} + \\
&\sum_{j=0}^{r(i)-2}
\left\lceil
\dfrac{
(r(i)-j-1) \cdot (r(i)-j)/2
}
{\#\mathrm{accelerators}}
\right\rceil
\cdot t_{gemm}^{(i)}
\end{align*}

\begin{figure}[t]
\centering
\includegraphics[width=0.65\linewidth]{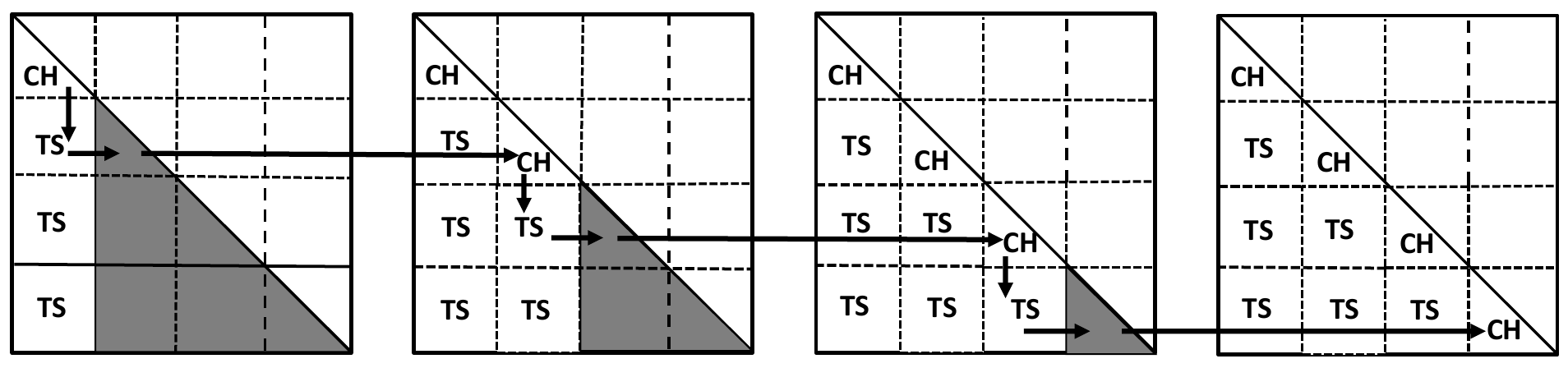}
\caption{Critical path for the blocked model of Dense Cholesky Factorization
for both the overlapping and the parallel (best-case) computation.
Dependencies are annotated with black arrows.
}
\vspace{-0.7cm}
\label{fig:blocked_cr_path}
\end{figure}

\begin{figure}[t]
\centering
\includegraphics[width=0.6\linewidth]{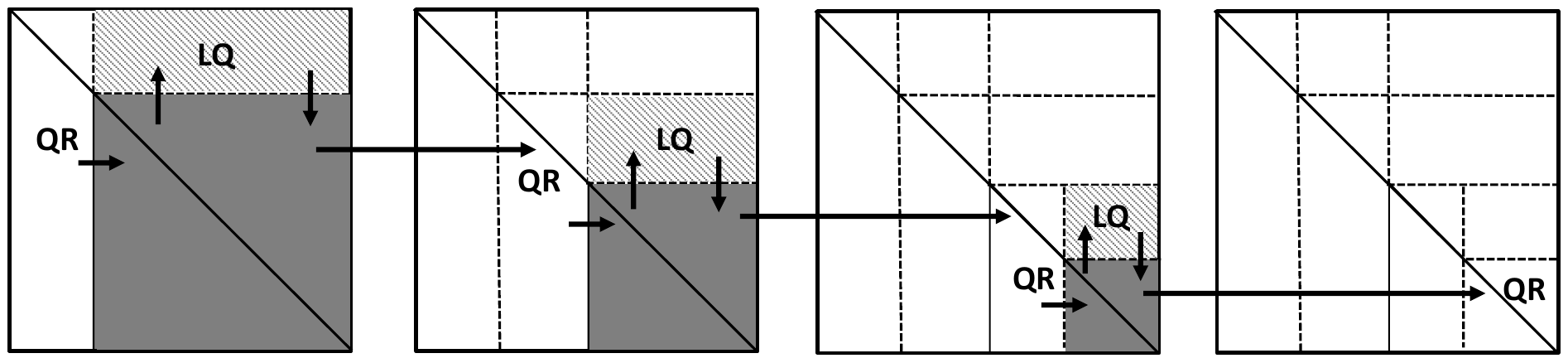}
\includegraphics[width=0.6\linewidth]{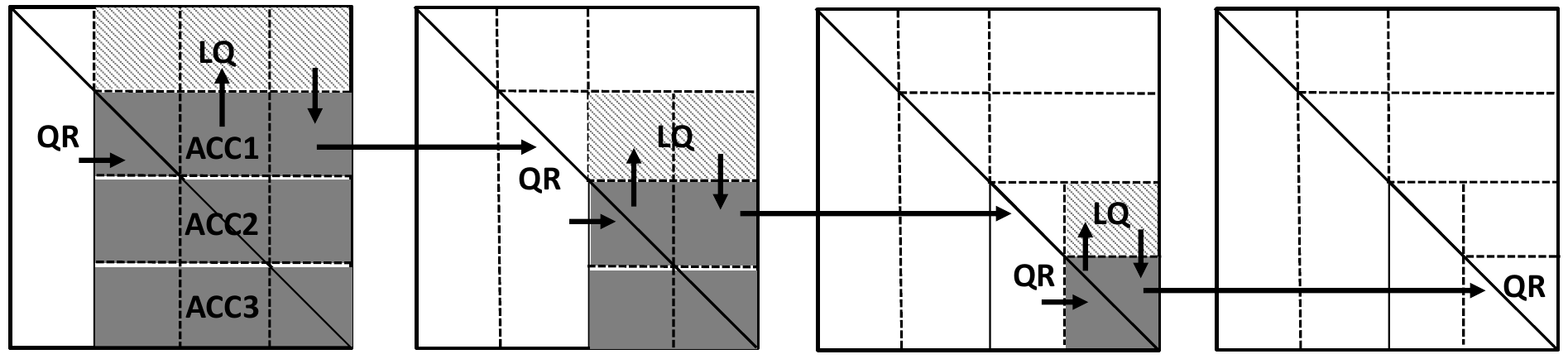}
\vspace{-0.3cm}
\caption{
The critical path of Iterative (top) and Blocked (bottom) models for 2-stage SVD (Refinement: 4).
Dependencies are annotated with black arrows.
}
\label{fig:svd-iter_cr_path}
\end{figure}
\subsubsection{\textbf{Overlapping/Pipeline Computation}}
An overlapping of the CPU and HW accelerator computation can occur, where there are no dependencies, thus minimizing the computation latency of the HW accelerated part using the blocked method.

The criteria for Cholesky are a) to fit $r(i)^2/8$ blocks
of gemm(i) computations in a HW accelerator and b) to ensure that the HW latency of the accelerator is less than the host side:
\begin{align*}
    \dfrac{r(i)^2}{8} \cdot t_{gemm}^{(i)} < min(t_{TS}^{(i)}, t_{CH}^{(i)})\  |\ i=0,\dots, N
\end{align*}

Given that both of the criteria above are satisfied, then the overall latency estimation is limited to the critical path of the blocked model, as depicted in Figure \ref{fig:blocked_cr_path}.

The model estimating the performance of the overlapping blocked approach (equivalent to the estimation of the latency of the critical path) is the following for Cholesky:
\begin{align*}
t_{comp}^{(i)} = r(i) \cdot t_{CH}^{(i)} + (r(i) -1) \cdot \left(t_{gemm}^{(i)} + t_{TS}^{(i)}\right).
\end{align*}

For SVD the blocked approach allows the use of multiple accelerators for simultaneous computation of gemm($i$) tasks. As shown in Figure~\ref{fig:svd-iter_cr_path}, a parallel implementation requires at least $r(2)-1=3$ accelerators to match the theoretical critical path. The performance model for the parallel blocked approach is given by:
\begin{align*}
t_{comp}^{(i)} = &\sum_{j=0}^{r(i)-1} t_{QR}^{(i,j)} + \sum_{j=0}^{r(i)-2} t_{LQ}^{(i,j)} + \left(r(i)-1\right) \cdot \dfrac{r(i)}{2} \cdot t_{gemm}^{(i)}.
\end{align*}

\vspace{-0.5cm}
\section{Experimental Setup}
\label{sec:setup}

Our methodology and computation models were evaluated using a setup emulating the hardware resources of a heterogeneous Supercloud. The hardware configuration includes a 48-core ARM CPU, specifically the Huawei Kunpeng~920~\cite{kunpeng}, paired with a Huawei Ascend~910 AI processor~\cite{ascend910}, which features 32 Da Vinci AI cores capable of reaching a peak performance of 320\,TFLOPS. Communication between these devices is facilitated via a PCIe link. For comparison, the CPU implementations in our experiments are based on multi-threaded, fully optimized versions of Dense Cholesky Factorization from LAPACK \cite{lapack}, \cite{dongarabooklapack1999}, as well as SVD functions and subroutines from LAPACK \cite{dongarabooklapack1999}.
The input for these experiments is a square matrix of size $32000$, resulting in $32000^2$ entries.

\section{Experimental Results}
\label{sec:exp}

\begin{figure}[t]
\centering
\includegraphics[width=0.5\linewidth]{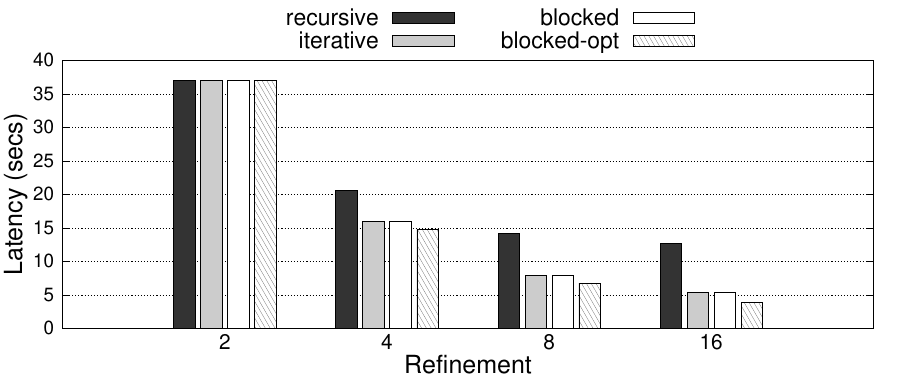}
\hfill
\hspace{-0.2cm}
\includegraphics[width=0.5 \linewidth]{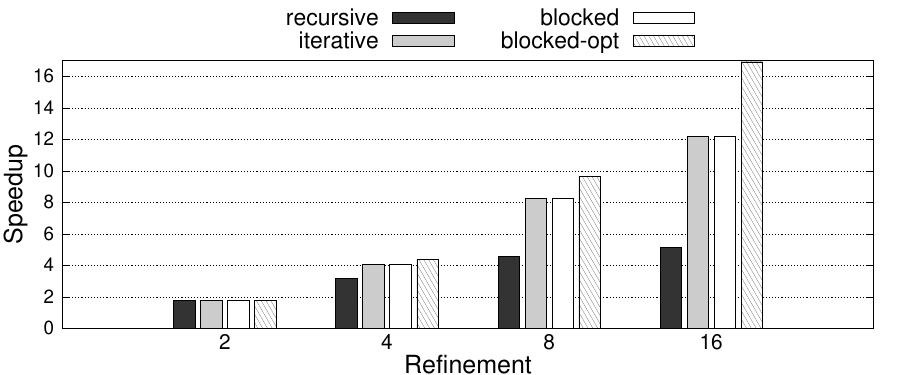}
\vspace{-0.7cm}
\caption {Latency (left) and Speedup (right) obtained for Dense Cholesky Factorization increasing the refinement level and gradually offloading more of the initial computation to the Ascend device. Comparison of the recursive, iterative, blocked and blocked-optimized (critical path) models.
}
\vspace{-0.5cm}
\label{fig:ch-speed}
\end{figure}

\subsection{Dense Cholesky Factorization -- Model Comparison}

The performance results for dense Cholesky (CH) Factorization allow a comparison among the three computation models explored. As seen in Figure \ref{fig:ch-speed} (Left), the total latency for the iterative and blocked approach is significantly decreased, compared to the recursive one, for refinement levels greater than two. This is due to larger amounts of gemm computations being accelerated and smaller part of the CH and TS computation residing at the CPU.

For comparison, the optimized (LAPACK \cite{lapack}), 48-core CPU-only  baseline latency
(refinement level=1)
is equal to 65 secs, whereas the blocked optimized version (refinement level=16),  where overlapping (critical path) is applied,  is equal to a mere 3.8 secs latency. We stop further refinement once the condition
$2 \cdot t_{CH}^{(i+1)} < t_{CH}^{(i)}\ |\ i=0,\dots, N$
, set in Section \ref{sec-rec}, is not satisfied.

Moreover, we notice that the blocked optimized version outperforms all other investigated models.
This is achieved by minimizing the latency both in the host-residing TS and the gemm-accelerated tasks. The latency of the CH tasks, on the other hand, is the bottleneck as they need to be executed sequentially due to existing dependencies.

Therefore, the performance of a blocked optimized strategy over an optimized, 48-core CPU-only baseline reaches a
striking \textbf{17x speedup gain}, as seen in Figure \ref{fig:ch-speed} (Right). Further refinement of the Cholesky application would not yield any more benefit as both the condition stated in Section \ref{sec-rec} is not satisfied and communication cost becomes more significant.

\vspace{-0.3cm}
\subsection{SVD}
\textbf{Bidiagonalization}
Our study examines the effects of refining task granularity and distribution, while offloading more computation to the first stage of the 2-stage bidiagonalization process.

\begin{figure}[t]
\centering
\includegraphics[width=0.5\linewidth]{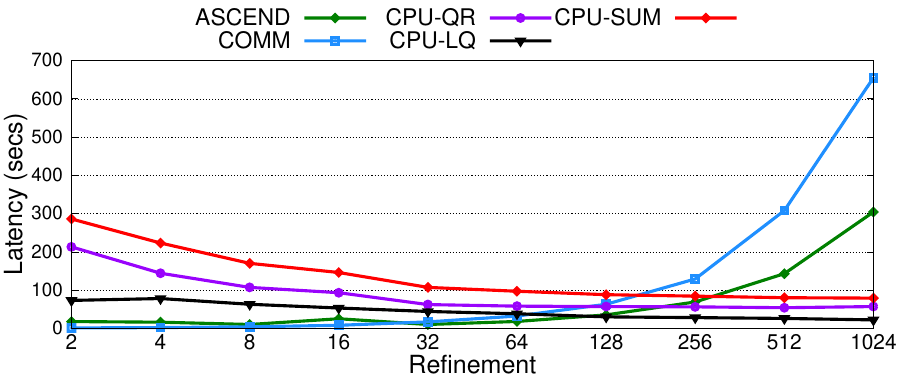}
\hfill
\hspace{-0.2cm}
\includegraphics[width=0.5\linewidth]
{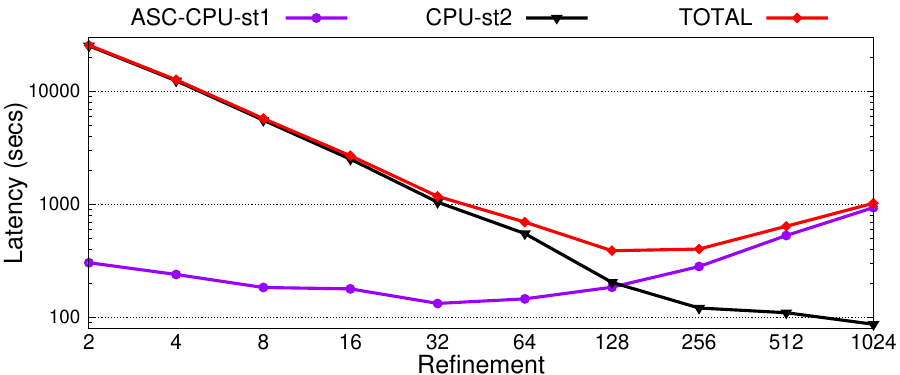}
\vspace{-0.7cm}
\caption {
Left: First stage of bidiagonalization (full-to-band): Latency of the gemm computation offloaded to the Ascend AI accelerator, communication between accelerator and host, and CPU time as the refinement increases. The total latency is approximately the maximum of 1) the total CPU time, 2) the Ascend time, 3) the communication time.
Right: Latency trade-off between the 1st and 2nd stages of bidiagonalization while varying the refinement. A refinement of 128 leads to minimal total latency.
}
\vspace{-0.5cm}
\label{fig:svd-stageone}
\end{figure}

Figure~\ref{fig:svd-stageone} (Left) illustrates the computation and communication dynamics between the CPU and the accelerator during the first bidiagonalization stage. We observe a gradual decrease in the host CPU's processing time, both for QR and LQ computation time.
On the other hand, the latency of the Ascend AI accelerator and communication overhead surpass the CPU time when refinement reaches levels 128 and 256. Beyond these levels, communication costs and accelerator latency increase exponentially, whereas CPU time plateaus.
These findings are corroborated by the results shown in Figure~\ref{fig:svd-stageone} (Right). As the refinement level progresses and more computational effort is offloaded to the AI accelerator and the host CPU, the second stage of bidiagonalization sees a reduction in CPU processing time. The combination of the two stages indicates that the optimal refinement levels are 128 and 256.

\begin{figure}[t]
\centering
 \includegraphics[width=0.5\linewidth]{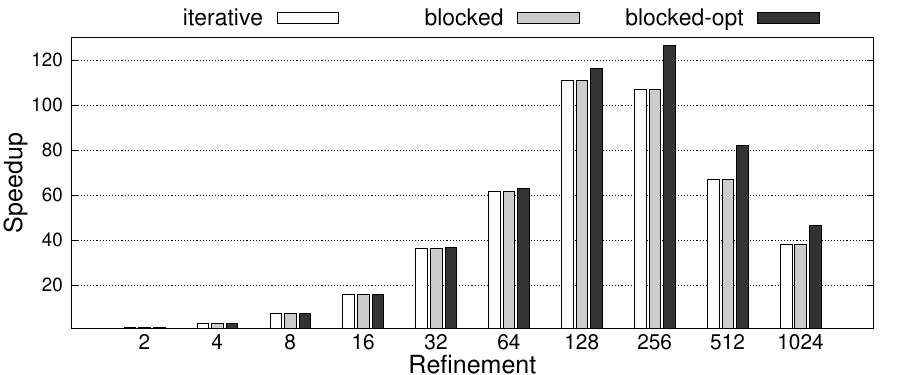}
\hfill
\hspace{-0.2cm}
\includegraphics[width=0.5\linewidth]{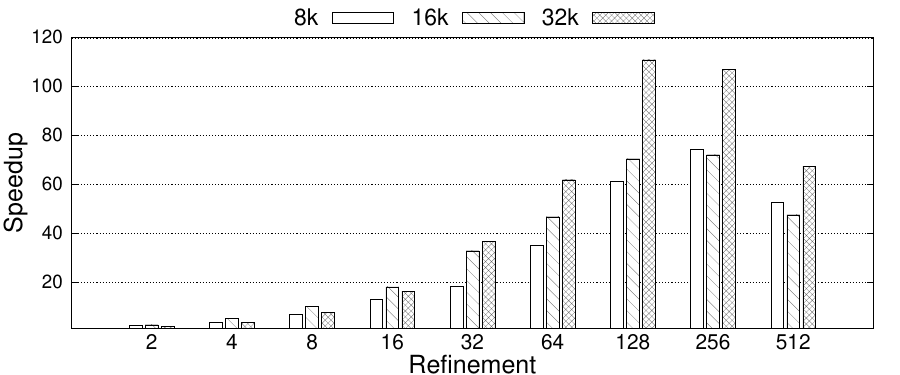}
\vspace{-0.6cm}
\caption{Left: Speedup obtained for SVD bidiagonalization across all three models of computation: iterative, blocked, and blocked-opt. (parallel). Gradually, as the level of refinement increases, we are offloading more computation to the Ascend AI accelerator.
Right: Speedup obtained for SVD bidiagonalization (iterative model) while increasing the refinement. Comparison among three input matrix sizes: 8k, 16k, 32k.}
\label{fig:svd-twostage-speed}
\vspace{-0.5cm}
\end{figure}

Figure~\ref{fig:svd-twostage-speed} (Left) confirms these findings by showcasing the speedup comparison of the overall bidiagonalization phase across the iterative, blocked, and blocked optimized models. Each model employs the Ascend AI accelerator(s) in conjunction with the host CPU, benchmarked against a fully optimized LAPACK 48-core ARM CPU version.
Regardless of the model used, we notice that the level of refinement is the \textbf{driving factor} of the performance achieved and, as expected by the previous graphs, speedup increases rapidly until the critical point of refinement 128 and 256 is reached.
The peak performance, with speedups of \textbf{116x} and \textbf{126x}, is observed at refinement levels 128 and 256 in the optimized parallel version of the blocked model.
Beyond these levels, performance declines sharply due to the escalating latencies of AI accelerators and communication costs, while CPU time remains stable.

The same trend regarding both peak speedup and refinement can be observed in Figure~\ref{fig:svd-twostage-speed} (Right), where we compare the speedup of the bidiagonalization phase across varying input matrix sizes of $8000$, $16000$ and $32000$ with $8000^2$, $16000^2$ and $32000^2$ entries respectively. As expected, speedup gains in absolute terms are greater for larger matrix sizes, compared to smaller ones. However the peak speedup points
are the same regardless of matrix sizes.\\

\begin{figure}[t]
\centering
\includegraphics[width=0.5\linewidth]{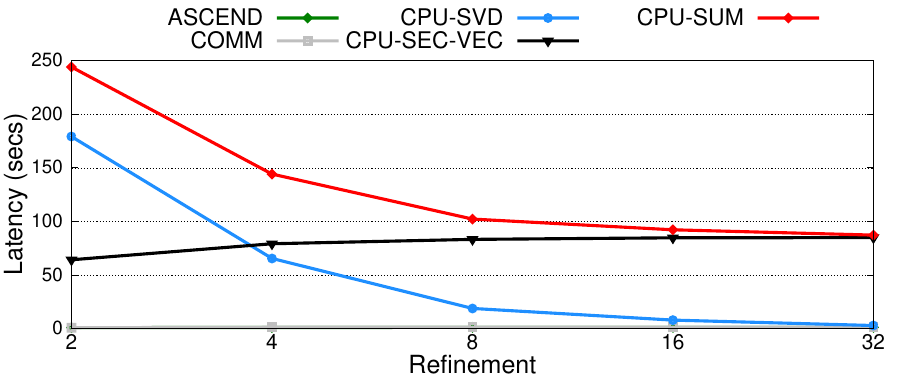}
\hfill
\hspace{-0.2cm}
\includegraphics[width=0.5\linewidth]{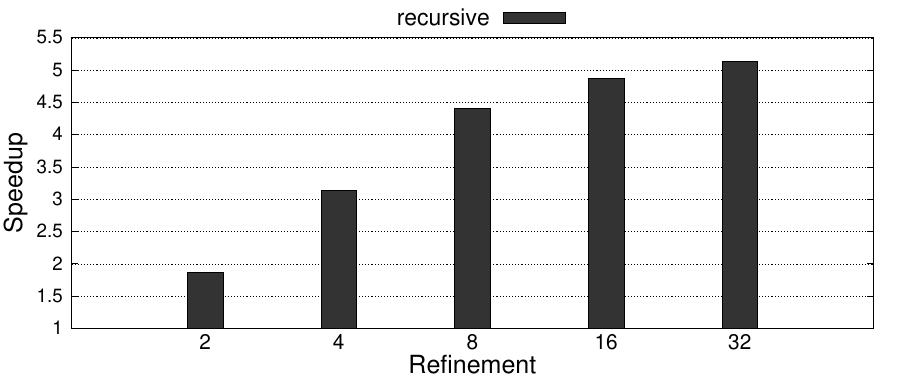}
\vspace{-0.7cm}
\caption{Left: Divide and Conquer latency of the Ascend AI accelerator, communication, and CPU time as refinement increases.
Right: Speedup obtained for SVD Divide and Conquer stage as refinement increases and employing the recursive model of computation.}
\vspace{-0.5cm}
\label{fig:svd-dc}
\end{figure}

\textbf{Divide and Conquer}
For the Divide and Conquer phase of SVD computation, the recursive model was employed. Figure~\ref{fig:svd-dc} (Left) displays the Ascend  AI accelerator latency, communication cost, and CPU latency (SVD computation and singular values/vectors computation) as refinement levels increase. Contrary to the bidiagonalization stage, as an increasingly larger part of the computation is gradually offloaded to the accelerators,
the accelerator latency and communication cost here do not significantly impact overall performance while refinement increases. The CPU time for SVD computation gradually decreases and stabilizes, while the time for computing singular values and vectors increases with refinement.

As most of the computation is being offloaded to the accelerator(s) for gemm tasks, the CPU time in SVD and singular values/vectors computation becomes critical for the performance of the Divide and Conquer phase. As shown in Figure~\ref{fig:svd-dc} (Right), the speedup improves as refinement increases, translating to more gemm task acceleration. It peaks at a 5.2x speedup with a refinement level of 32, whereafter CPU computation for SVD plateaus.

It can be observed that in this phase, refinement cannot be extended more than five iterations (Iteration: $i=5$, Refinement: $r(i)=2^5=32$) to gain more in performance, however the speedup achieved can still contribute significantly to the overall SVD computation.\\

\textbf{Back-Transformation}

\begin{figure}[t]
\centering
\includegraphics[width=0.5\linewidth]{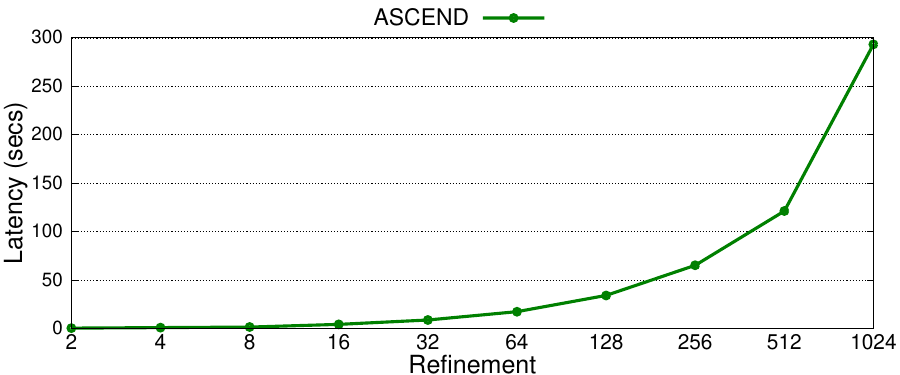}
\hfill
\hspace{-0.2cm}
\includegraphics[width=0.5\linewidth]{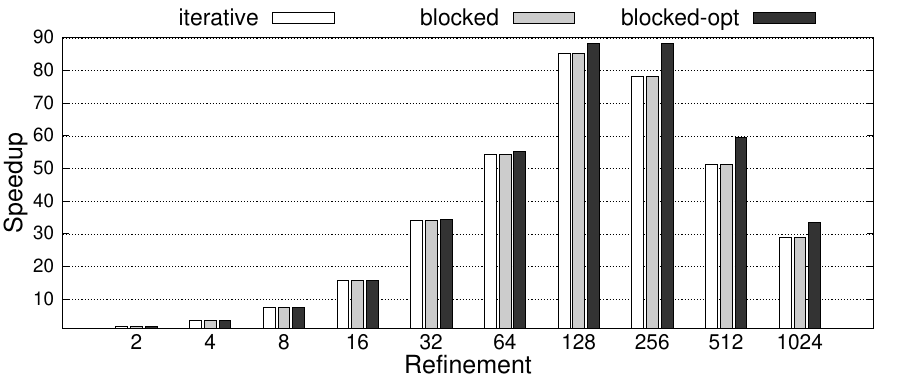}
\vspace{-0.7cm}
\caption {Left: Back-transformation. Ascend AI accelerator latency as refinement increases.
Right: Speedup obtained for the entire computation of SVD (bidiagonalization, Divide and Conquer, and back-transformation) increasing the refinement. Comparison of the iterative, blocked, and blocked-opt. (parallel) models.}
\vspace{-0.5cm}
\label{fig:svd-bt}
\end{figure}

The final SVD computation stage involves a three-way back-transformation to extract singular vectors. This stage relies solely on accelerators for matrix-matrix multiplications without a specific computation model. In Figure~\ref{fig:svd-bt} (Left), the latency of the AI accelerator is shown to initially increase and then spike exponentially at refinement levels between 128 and 1024. This escalation results from increasingly larger gemm tasks being offloaded at each iteration.

As discussed in Section~\ref{sec:bt}, the full-to-band reduction's computational demand directly influences the first back-transformation's load, impacting overall SVD performance in a heterogeneous Supercloud system.
Therefore, the more computational effort takes place in the full-to-band reduction stage, the more computation is required in the first back-transformation stage. This is an important factor and one that can severely limit overall SVD computation performance.

\vspace{0.3cm}
\textbf{End-to-End SVD Computation}
The analysis of all three SVD stages reveals insights into the overall performance of SVD applications in a heterogeneous Supercloud. Figure~\ref{fig:svd-bt} (Right) compares the speedup between a Supercloud setup and a homogeneous multi-core CPU for an end-to-end SVD computation.

The comparison evaluates all stages of computation, with most tasks offloaded to AI accelerators, against a fully optimized CPU-only setup. In the bidiagonalization stage, we assess iterative, blocked, and blocked optimized models. The recursive model is utilized for the Divide and Conquer stage, while back-transformation computations are incrementally offloaded to the Ascend AI accelerator.
The Supercloud achieves a peak speedup of 85x for iterative and blocked models at a refinement level of 128, and an 88x speedup for the blocked optimized model, where parallelism is maximized, at refinement levels 128 and 256. Notably, for refinement levels below 128, the choice of computation model does not significantly influence the overall speedup.

\section{Conclusions}
We introduce novel  models of performance that we developed and are based on the recursive, iterative, and blocked computation models. Leveraging our tool-chain, we have managed to automatically map Dense Cholesky (CH) Factorization and Singular Value Decomposition (SVD) onto a heterogeneous system, consisting of a Kunpeng 48-core ARM CPU and an Ascend AI accelerator device, achieving up to a 17x speedup for CH and up to 88x speedup across the entire end-to-end SVD computation, over fully-optimized ARM CPU baseline implementations.

\bibliographystyle{unsrt}
\bibliography{references}

\end{document}